# *Operando* Plasma-XPS for Process Monitoring: Hydrogenation of Copper Oxide Confined Under h-BN Case Study

J. Trey Diulus,[a]* Andrew E. Naclerio,[b] Anibal Boscoboinik,[c] Ashley R. Head,[c] Evgheni Strelcov,[a] Piran R. Kidambi,[b,d,e] and Andrei Kolmakov[a]*

[a.] Nanoscale Device Characterization Division, PML, NIST, Gaithersburg, MD, 20899, USA

[b.] Department of Chemical and Biomolecular Engineering, Vanderbilt University, Nashville, TN 37212, USA

[c.] Center for Functional Nanomaterials, Brookhaven National Laboratory, Upton, NY 11973

[d.] Department of Mechanical Engineering, Vanderbilt University, Nashville, TN 37212, USA

[e.] Vanderbilt Institute of Nanoscale Sciences and Engineering, Nashville, TN 37212, USA

*john.diulus@nist.gov; andrei.kolmakov@nist.gov

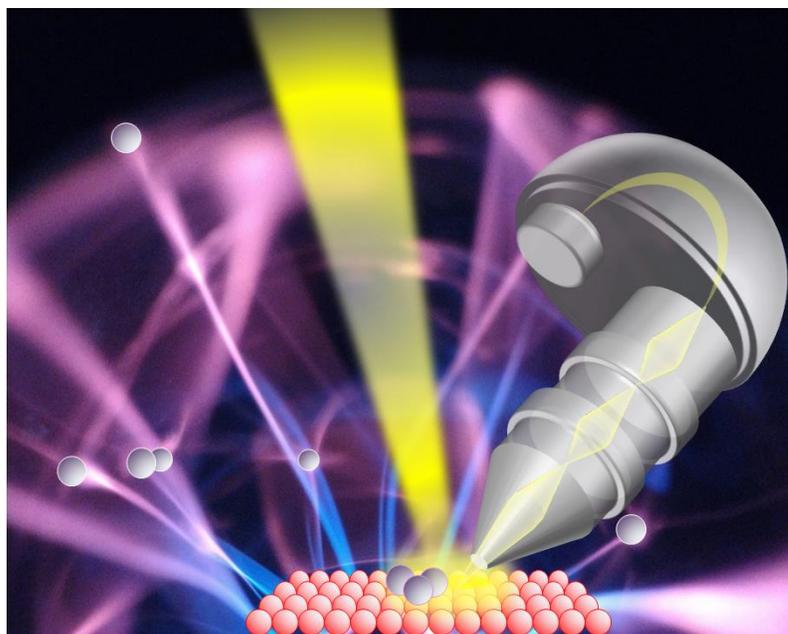

TOC figure

## Abstract

We demonstrate that ambient pressure x-ray photoelectron spectroscopy (APXPS) can be used for *in situ* studies of dynamic changes in surface chemistry in a plasma environment. Hexagonal boron nitride (h-BN) was used in this study as a model system since it exhibits a wide array of unique chemical, optical and electrical properties that make it a prospective material for advanced electronics. To better understand the stability and surface chemistry of h-BN during plasma assisted processing, we used polycrystalline Cu foils with single layer h-BN, grown via chemical vapor deposition (CVD), and track in real time the plasma induced reduction of the underlying Cu oxide using APXPS equipped with 22 kHz 75 W discharge plasma source operating at 13 Pa. Residual gas analysis (RGA) mass-spectra were concurrently collected during plasma-XPS to track reaction products formed during plasma exposure. A clear reduction of $Cu_xO$ is seen, while an h-BN layer remains intact, suggesting H radical (H•) species can attack the exposed and h-BN covered Cu oxide patches and partially reduce the underlying substrate. In addition to demonstration and discussion of plasma-XPS capabilities, our results indicate the h-BN encapsulated metallic Cu interface might be repaired without significantly damaging the overlaying h-BN, which is of practical importance for development of h-BN encapsulated devices and interfaces.

## I. Introduction

Modern ambient pressure x-ray photoelectron spectroscopy (APXPS) measurements are conducted at pressures spanning approximately $10^{-10}$ Pa to $10^3$ Pa range, allowing for a robust metrology technique of choice for *in situ* studies of realistic surfaces and interfaces, specifically for gas-solid or gas-liquid interfacial reactions (see recent reviews[1-3] and references therein). There is a clear recent trend in application of APXPS to *operando* studies relevant to semiconductor microfabrication technologies.[4] In this regard, most cold plasma semiconductor processing occurs at $10^{-1}$ Pa to $10^3$ Pa, which overlaps with the operational pressure range for APXPS. Therefore, application of APXPS to interrogate the surfaces and interfaces in a plasma environment is, in principle, possible and would be greatly beneficial for plasma assisted process monitoring in semiconductor fabrication technology, where the real time analysis of interfacial chemical composition with sub monolayer precision is a common requirement. However, no systematic efforts have been undertaken, to the best of our knowledge, until recently,[5] to collect XPS spectra under plasma operating conditions. Instead, a traditional *ex situ* "before-and-after" approach is usually employed.[6] Here, we utilize polycrystalline Cu foils with a hexagonal boron nitride (h-BN) monolayer grown via chemical vapor deposition (CVD) to demonstrate plasma-XPS capabilities for studying plasma induced surface chemistry and confined reactions.

h-BN has garnered extreme interest in the advanced electronics community due to a wide array of unique chemical, thermal, and electrical properties that presents itself useful in numerous optical and electronic applications (see reviews ref. [7-9] and references therein). h-BN capped heterostructures and interfaces have been of interest for solid state lighting, high power/temperature electronics etc. Due to a wide bandgap of about ~6 eV, chemical inertness, and breakdown strength, h-BN has also been suggested as a prospective traps-free gate dielectric material, showing particularly impressive performance in 2D electronics and high mobility diamond transistors.[10] Multiple studies have shown that h-BN can further act as a passivation layer, to resist complete oxidation of an underlying substrate[11] or impede interlayer diffusion through stacked materials in electronic devices.[12] Cu is an ideal model substrate material for testing boron nitride as an encapsulant because h-BN can be epitaxially grown via CVD on any Cu orientation, with a near perfect lattice match achievable on the Cu(111) surface.[7] Furthermore, the synthesis/transfer of h-BN/Cu to arbitrary substrates can routinely be done using variety of wet or dry methods.[13-14]

An ideal CVD grown defect-free single crystal h-BN layer would completely prevent oxidation of the Cu substrate, however, intercalation of $O_2$ through the point/linear defects in an h-BN monolayer can still occur in realistic samples, yielding an underneath Cu surface predominantly oxidized to copper(I) oxide ($Cu_2O$) instead of complete oxidation to copper(II) oxide (CuO).[11] This intercalative oxidation can occur thermodynamically, even at room temperature if the partial pressure of oxidizing species is high enough, limiting a lifetime of pristine unoxidized h-BN/Cu heterostructures to a few weeks in an ambient environment.[15-17] In addition to intercalative oxidation, deintercalation has also been demonstrated,[15] although the ability to subsequently intercalate a reducing molecule and recover the initial interface without destroying the h-BN layer has been more challenging to achieve.

Transport through 2D membranes is well-known, and has been suggested as a method for improvement across several electronic applications.[18] Prior $H_2$ exposure experiments to intercalate molecular hydrogen under h-BN showed that the process can occur either through the defects or via intercalation at h-BN/Cu interface due to lower energy barrier for molecular dissociation and diffusion at the interface.[19] On the other hand, plasma generated hydrogen radicals (H•), have an atomic diameter smaller than the h-BN lattice constant and can potentially penetrate directly through the hexagonal ring, thus increasing the probability of reacting with the confined oxide. A previous report[20] shows not only can remote plasma H• radicals intercalate through h-BN, but form hydrogen "bubbles" after intercalation due to recombination of radicals underneath the h-BN. Interestingly, exposure to an Ar and $O_2$ plasma do not show such behavior supporting the radial/ion size selectivity of the intercalation process. The reaction and modification of the h-BN itself due to reaction with hydrogen radicals has been reported as well.[21-22]

Overall, understanding plasma-induced surface (intercalation) reactions is still elusive and would greatly benefit from *operando* plasma-XPS capabilities that are demonstrated and discussed in this report. Using polycrystalline Cu foils with CVD grown h-BN that have been slightly oxidized in ambient as a model system, we comparatively assessed the reduction of the surface via two hydrogenation methods: (i) simple exposure to $H_2$ at 13 Pa followed by annealing in vacuum (sequentially) and annealing during $H_2$ exposure (simultaneously), and (ii) (ii) exposure to a low power (75 W) $H_2$ remote plasma also at 13 Pa with APXPS. Complementary scanning electron microscopy (SEM) provides morphological maps of foil surfaces before and after hydrogen exposure. Through use of plasma-XPS, we demonstrate the ability to track dynamic changes related to plasma chemistry. Ultimately, we show the H• can interact with the confined oxide and reduce the surface, suggesting a route to reduce the Cu substrate and recover the original h-BN/Cu interface is possible within a certain parameter space.

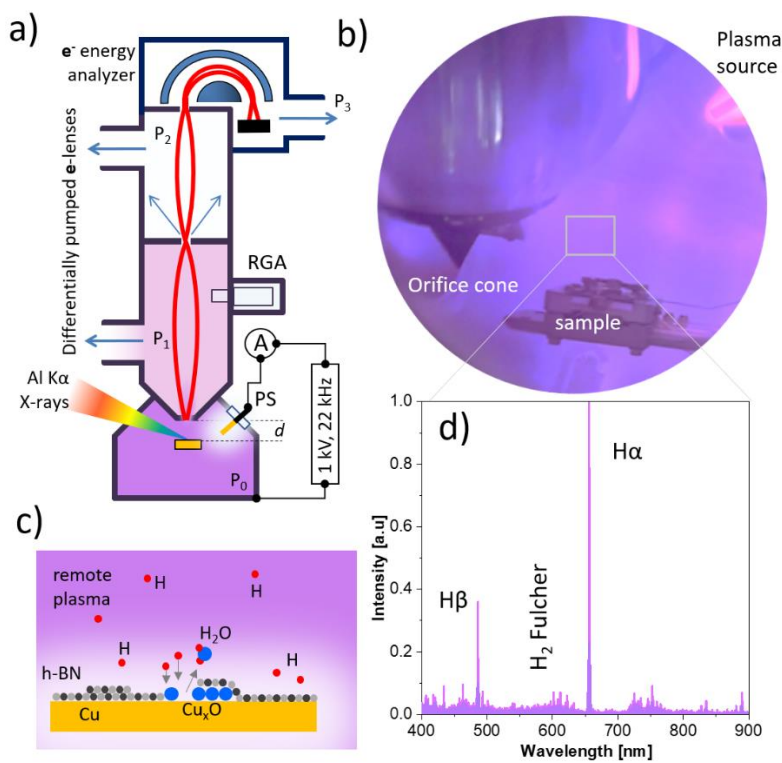

**Figure 1**. Schematic of BNL APXPS system with reaction chamber, plasma source (PS) and residual gas analyzer (RGA) (a); Photograph of

## II. Experimental setup

Single monolayer h-BN was grown on polycrystalline Cu foil via chemical vapor deposition using a high-pressure/high-temperature reactor setup, described in Naclerio et al.[7] The h-BN/Cu samples were briefly exposed to atmosphere and then stored in vacuum (ca. ~10 Pa) inside a desiccator box for approximately 1 year. We used these slightly oxidized h-BN foils as a model system for studying the recovery of the original metallic Cu and h-BN interface. Thus, the degree of oxidation across all samples slightly varies, although all samples were cut from the same original 6 cm × 2 cm source foil.

Plasma-XPS experiments were performed at the Center for Functional Materials at Brookhaven National Laboratory equipped with lab-based APXPS system (Figure 1a).[23] The AP-XPS instrument utilizes a reaction chamber with a base pressure of <5 × 10$^{-7}$ Pa and the ability to backfill the entire chamber with hydrogen gas during data collection. The reaction chamber is separated from the multi-stage differentially pumped electrostatic focusing lens system and electron spectrometer by a 300 μm diameter cone aperture to enable XPS data collection at pressures up to ca. 200 Pa (Figure 1a,b). The sample position was adjusted to the focal point of the X-rays (600 um below the aperture) by optimizing the intensity of a photoemission peak. A monochromatized Al Kα X-ray source (hv = 1,486.6 eV), focused to a ca. 300 μm diameter spot size and fixed at 55° from the sample normal, was used for acquiring XPS spectra. Survey spectra were collected using a pass energy of 50 eV with a dwell time of 100 ms and a step size of 1 eV. High resolution spectra were collected with a 20 eV pass energy, 250 ms dwell time, 50 meV step size, and sufficient sweeps for decent signal-to-noise, with doubled number of sweeps for data collected at elevated pressure. A standard set of scans would consist of a survey, Cu 2p, O 1s, N 1s, C 1s, B 1s, and valence band, which would total roughly 1 hour of scan time. Sample heating was performed by controlling the power of an infrared laser aimed at the backside of the sample holder with an abrasive coating to increase the absorption and diffusivity of the laser for optimized heating. The temperature was monitored *in situ* with a K-type thermocouple spot-welded to the sample plate.

The first APXPS differential pumping stage is equipped with quadrupole mass spectrometer for residual gas analysis of the reaction chamber environment during ca. 100 Pa of hydrogen exposure (Figure 1a). The secondary electron multiplier was used for high signal-to-noise residual gas analysis (RGA) scans, as the pressure in the first stage was below 10$^{-4}$ Pa. A survey analog scan was first collected for a mass to charge ratio (m/z) of 1 Da/e to 100 Da/e of the background ultra-high vacuum (UHV), and then during chamber backfilling of 13 Pa H$_2$, and during the ignited plasma at 13 Pa H$_2$, all collected with a sweep time of ca. 33.5 seconds. During each APXPS spectrum scan, with and without ignited plasma, RGA data were the ignited plasma at 13 Pa (b); Diagram of plasma surface interaction for h-BN/Cu (c); Remote H$_2$ plasma emission spectrum collected near the sample also at 13 Pa (d).

collected only for specific masses (m/z = [2, 16, 18, 28, 32, 40, and 44] Da/e) for increased time resolution, using 100 ms dwell time for each m/z. To differentiate the on-sample reactions from the side (on-walls) reactions, RGA was collected where no sample was present in front of the cone-orifice during plasma.

Plasma was generated in the APXPS chamber by backfilling the analysis chamber to 13 Pa of $H_2$ and applying 1 keV peak-to-peak AC voltage (22 kHz) to in-chamber copper electrode through a high voltage feedthrough. The power supply was set to 12% output for all plasma exposures. The copper electrode was located 10 cm from the sample during XPS spectra acquisition well beyond discharge area (Figure 1b). Therefore, the sample is under remote plasma conditions where the concentration of ions and electrons is low, and surface redox reactions are induced mainly by hydrogen radical species (Figure 1c). This is confirmed by a hydrogen plasma afterglow spectra that exhibits a violet hue, seen in Figure 1b, due to the mixing of H-alpha ($H_\alpha$) and H-beta ($H_\beta$) visible spectral lines of the hydrogen atom at 656 nm and 486 nm, respectively. Optical emission spectrum of hydrogen plasma is also shown in Figure 1d, where the major atomic recombination peaks $H_\alpha$ and $H_\beta$ are seen, in addition to the lower intensity broad molecular band ($H_2$ Fulcher) generated from electron induced fluorescence of the $H_2$ molecule.

Complementing SEM measurements were conducted in a separate UHV SEM/SAM system, at a base pressure of $<1 \times 10^{-7}$ Pa. The typical electron beam parameters were: probe current within 100 pA to 2 nA range; beam energy 3 kV; and field of view of ca. 100 $nm^2$ to 100 $\mu m^2$. The sample was angled at 15° to collect slightly grazing emission images and enhance the contrast at the surface. SEM Images were collected using an in-lens detector, which is sensitive predominately to low energy secondary electrons. The UHV SEM/SAM system also has an XPS chamber for traditional UHV data collection, using a monochromatized Al K$\alpha$ X-ray source (hv = 1,486.6 eV) focused to 1.5 mm × 2 mm oval spot size and angled 30° from the sample normal. The electron energy analyzer is positioned for 30° electron emission (thus, 90° angle from source to analyzer) and set to a 6.3 mm lens aperture with a 3.3 mm × 11 mm exit slit opening. All survey spectra were collected using 100 eV pass energy, 200 ms dwell time, 0.5 eV step size, and 2 sweeps. For high resolution spectra, the pass energy was lowered to 20 eV, with a step size of 50 meV, a dwell time of 300 ms, and sufficient sweeps to achieve a decent signal-to-noise. All XPS spectra were peak fit using a Shirley background and normalized to the lower binding energy baseline to account for occasional changes in X-ray source intensity.

## III. Results and discussion

The data acquisition flow was the following: after collecting baseline spectra and SEM images of the as-received sample, shown in red in Figure 2a-d and Figure **2**e, respectively, we backfilled the chamber with 13 Pa of $H_2$ and collected spectra continuously during exposure to molecular hydrogen, leading to an exposure time of roughly one hour. At room temperature, no significant redox chemistry is expected to take place outside of potential intercalation of $H_2$. Cu 2p spectra in Figure 2a collected before (red) and after (green) exposure to 13 Pa $H_2$ show Cu(I) satellite features at BE ca. 944 eV, in addition to three peak fits of the $2p_{3/2}$ peak, corresponding to metallic $Cu^0$ (932.6 eV, red), $Cu^+$ copper(I) oxide (933.7 eV, green), and $Cu^+$ copper(I) hydroxide (935.4 eV, blue). This agrees well with before-and-after O 1s spectra shown in Figure 2b, where one primary peak corresponding to O-Cu (530.5 eV, red) is seen along with adventitious species corresponding to O-C or O-H (532.0 eV, green) and adsorbed water (533.2 eV, blue). Thus, room temperature exposure to molecular $H_2$ at 13 Pa shows no evidence of reduction of the Cu substrate, with the only difference being a slight decrease in the adsorbed $H_2O$ peak relative to the O-C/O-H peak, which can be explained by a change in vapor equilibrium with the introduction of $H_2$ gas. Similarly, desorption of water (and potentially other volatile adventitious species) by introducing $H_2$ explains the increase in total signal of all peaks relative to the background.

Similarly, virtually no change is seen in the N 1s (Figure 2c) before and after exposure to molecular $H_2$, where two peaks can be fit corresponding to N-B (398.1 eV, red) and N* (399.1 eV, green). The shoulder peak (N*) arises from defects in the BN layer that can be attributed to domain boundaries, multilayers, and/or differences in h-BN/Cu interaction where a stronger interaction leads to charge transfer and creates a shift to higher binding energy.[15, 24] Intercalation of molecules like CO can cause the XPS binding energy to shift even further as the h-BN/substrate interaction changes due to intercalation.[24] This phenomenon is also present for the B 1s and can explain the change seen in Figure 2d, where initially single peak fit corresponding to B-N (190.5 ev) then develops a shoulder peak seen at 191.5 eV after $H_2$ exposure. In this case, exposure to, and intercalation of, $H_2$ can be weakening the interaction of B-Cu, or additionally $H_2$ intercalation between multilayer islands (triangles seen in Figure 2e, f) can create a similar peak shift. In our case, we use only one additional peak fit to account for these peak shifts in the N 1s and B 1s, although there are likely multiple peaks with similar energies that create this shoulder feature.

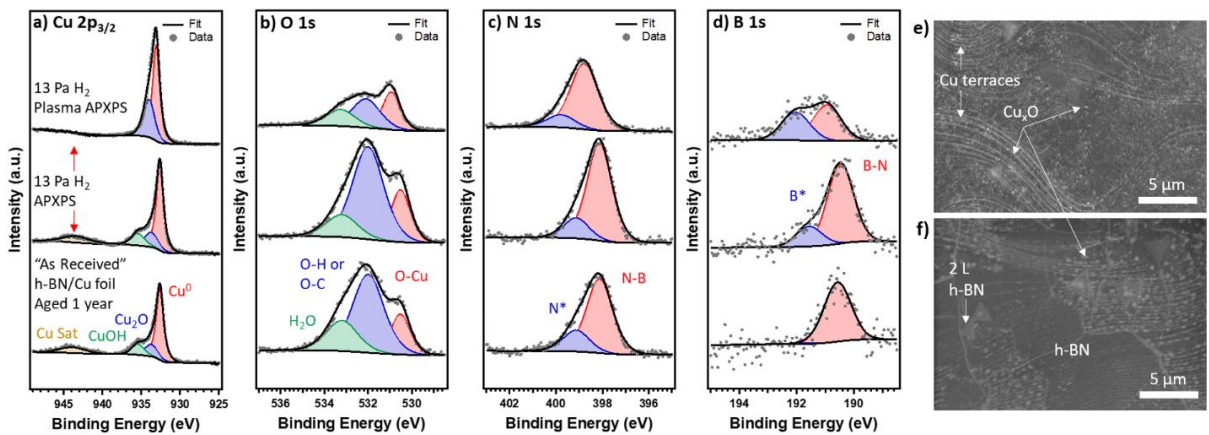

**Figure 2**. XPS collected for Cu 2p (a), O 1s (b), N 1s (c), and B 1s (d) of "as-received" (red) h-BN/Cu foils, followed by APXPS in 13 Pa partial pressure of $H_2$ (green), and finally during plasma exposure also in 13 Pa partial pressure of $H_2$ (blue), all collected at room temperature. Typical SEM images the h-BN/Cu stack collected before (e) and after plasma exposure (f) using 2 nA and 1.5 nA beam current, respectively, with the same energy (3kV) and detector settings.

Following molecular exposures, we ignite a cold plasma described in the experimental section where the discharge region was positioned 10 cm away from the sample to expose the sample predominantly to hydrogen radicals (remote plasma conditions). Preliminary experiments (not shown) indicated that direct plasma exposure can cause the h-BN single layer to be etched away within a few minutes. Instead, with a remote plasma, due to longer inelastic mean free paths of radical species, the primary species interacting with the sample will be hydrogen radicals, opposite to electrons and ions under direct plasma exposure that etch h-BN.

In Figure 2a, a clear loss in Cu 2p satellites takes place during *operando* spectra acquisition under plasma exposure, in addition to the removal of the CuOH shoulder, signifying a clear reduction of the Cu although some $Cu^+$ component remains. The O 1s in Figure 2b shows a decrease in the relative ratios between the adventitious species and the oxidized Cu from the substrate, confirming the plasma can also interact with and "clean" the surface, although these adventitious species can be redeposited to the surface rather quickly at this pressure upon the quenching of the plasma. Still the N 1s remains mostly the same, although the B 1s shows a significant increase in the secondary higher energy peak, again suggesting that intercalative species may be affecting the interaction between B and the Cu substrate. While it is possible that this increase in B* can be from newly formed defects in the BN lattice, however, then we should see a similar increase in the N* component of the N 1s spectra if BN bonds break and are replaced by either B-H bonds as a result of reaction with plasma radicals or B-O from reaction with the $Cu_xO$ surface.

A noticeable peak shift of ca. 0.5 eV to higher binding energy does occur during exposure to plasma. We tentatively attribute this to be an electronic effect of local work function increase due to the presence of non-zero plasma potential, since other measured core-level binding energies were referenced to the vacuum level in UHV (or inert gas) conditions. Changes in binding energies have been reported for adsorbed C with respect to the vacuum level,[25] and also for intercalated species, as for B 1s and N 1s of h-BN due to weakening interaction with the substate.[24] However, the energy shift seen in our case is to higher binding energy and a peak shift occurs also in the Cu 2p spectra, which should still be grounded to the spectrometer and thus unaffected by changes in the vacuum level. Thus, we presume that this higher binding energy shift occurs due to the presence of potential drop within weak plasma sheath that is formed even under remote plasma conditions, and which is more pronounced in XPS BE when using an insulating sample such as thicker h-BN or bulk diamond (not shown here).

SEM images were also collected *ex situ* before and after exposure to plasma presented in Figure 2e and Figure **2**f, respectively. Prior to exposure, as received slightly oxidized h-BN/Cu sample surface is covered with carbonaceous (darker patches) and oxidized areas (bright spots) appeared on otherwise smooth h-BN/Cu gray background. Oxidized spots often decorate Cu staircasing valley edges. Some 2× multilayer h-BN triangles can also be observed. After plasma exposure, the fraction of the pristine (gray) h-BN/Cu area noticeably increases with concomitant coalescence of the oxidized (bright) clusters and overall decrease of O1s peak intensity. Simple particle analysis of the before-and-after SEM images of the oxidized patches corroborates with XPS observations confirming that the plasma is reacting with the C/O species and "cleaning"/reducing the h-BN/Cu surface, while the BN seems to remain intact.

We now discuss the necessity and also one of the challenges of in-chamber plasma-XPS. To get a sense of potential reactions occurring on the sample, we used the RGA connected to the first APXPS pumping stage, which utilizes the cone orifice as a "sniffing" probe of the plasma induced reaction products originating at the h-BN/Cu sample surface. Since the sample-to-orifice distance was in the same order of magnitude as average molecular mean free path at 13 Pa of hydrogen (standard plasma conditions), the assumption was that a significant fraction of the molecules entering the cone orifice would experience at least a single collision with the sample surface and would be related to the products of plasma induced surface reactions. To test the feasibility of this approach and to discriminate between possible artifacts, we collected comparatively RGA spectra during running plasma conditions when a sample was (i) moved far away from the optimal "sniffing" conditions and (ii) in front of the cone-orifice.

Figure 3 shows temporal evolutions of ion currents for few major molecular components for 5 min exposure to plasma (a), and during the plasma APXPS (b) at RT. During the 5 min exposure in Figure 3a, the sample was moved away from the cone, and is majority defined by the changes in gas composition inside the analytical chamber upon plasma application. Initially, as the $H_2$ is backfilled into the chamber we see an uptake of all masses, which is common for APXPS as the increase of the background pressure leads to outgassing from the chamber walls.[26] As the plasma is ignited, a noticeable immediate increase in the O (m/z 16 Da/e, blue), CO (m/z 28 Da/e, aqua) and $CO_2$ (m/z 44 Da/e, red) signals can be observed directly corresponding with the plasma onset. A slight increase in the $H_2O$ (m/z 18 Da/e, gray) is also seen. No significant changes can be seen for $H_2$ (m/z 2 Da/e, gold). We interpret the observed results as plasma induced reactions with water and hydrocarbon molecules adsorbed at the chamber walls. Note, qualitatively similar responses for all the same molecular components can be observed during the APXPS measurements under plasma conditions when the sample is in front of and in proximity to the "sniffing" cone (Figure 3b). This fact manifests two challenges: (i) the crucial and potentially parasitic role of the side-wall reactions on to the surface chemistry of the sample of interest and (ii) the need in optimization of sample-selective mass spectroscopy in global plasma environment. Some of the potential solutions are proposed in the next section.

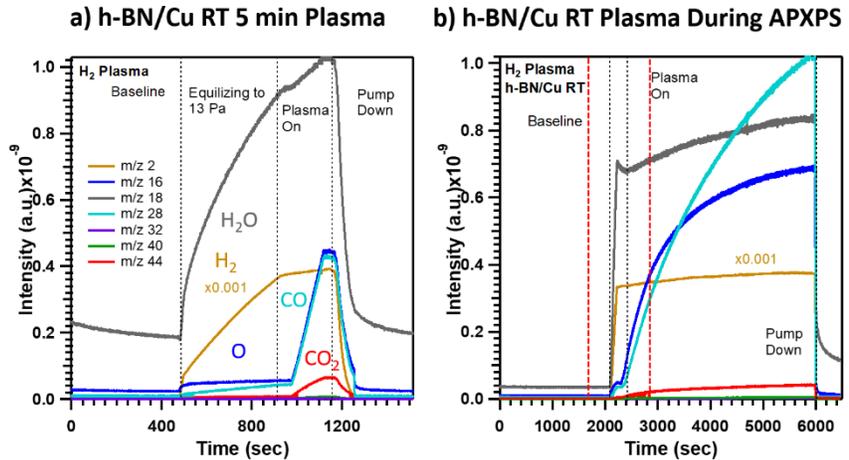

Figure 3. RGA spectra collected during 5 min hydrogen remote plasma exposure with the sample moved away from the "sniffing" cone (a), and during plasma-XPS (b). $H_2$ signal (m/z = 2 Da/e, gold) is reduced by a factor of $10^{-3}$ for scaling.

## IV. Conclusions and outlook

To summarize, we demonstrated that good quality XPS spectra can be collected during running plasma conditions in the analysis chamber. Using CVD grown h-BN monolayer Cu foil, we were able to monitor the reduction of the oxidized $Cu_xO$ patches under remote plasma conditions, without the destruction of the h-BN overlayer. We indicated that the measured core level energy can experience the shift tentatively assigned to the presence of a non-zero plasma potential near the surface under plasma conditions. This phenomenon will be even more pronounced when the sample is immersed into the plasma discharge environment and is still awaiting further experiments. We also found that care must be taken in interpreting the plasma induced reactions at the sample surface when global plasma in the analytical chamber induces numerous parasitic reactions at the chamber walls that, in turn, can alter the sample surface. The potential solutions of the problem would be localization of the plasma using plasma micro-jets or employing small size sample enclosures with inert walls and micro-plasma discharge inside. Such a hardware, in principle, is already available at some APXPS setups and requires minor customization for (micro-) plasma source incorporation.[2, 27, 28]

Overall, plasma-XPS approach is an emerging new metrology platform that will greatly complement plasma chemistry as well as plasma-surface interactions fundamental research and is currently particularly demanded for process monitoring in semiconductor industry, bio-medical plasma applications and plasma remediation technologies.

# V. Associated Content

## Author Contributions

The manuscript was written through contributions of all authors. All authors have given approval to the final version of the manuscript.

Conceptualization: JTD, AEN, AB, AH, ES, PK, AK

Methodology: JTD, AEN, AB, AH, ES, PK, AK

Investigation: JTD, AB, AK

Visualization: JTD, AK

Funding acquisition: JTD, AB, AH, PK, AK

Project administration: AK

Supervision: AK

Writing – original draft: JTD, AK

Writing – review & editing: ES, ZN, NC, AEN, PK, JO

## Data And Materials Availability

The data both presented and not presented in this report, which support the findings of this study, are available upon reasonable request of the authors.

## Conflicts of interest

There are no conflicts to declare.

## Acknowledgements/Funding

We would like to acknowledge Glenn Holland and David Rutter (both at NIST) for technical support. This research used the Proximal Probes Facility of the Center for Functional Nanomaterials (CFN), which is a U.S. Department of Energy Office of Science User Facility, at Brookhaven National Laboratory under Contract No. DE-SC0012704. P.R.K. acknowledges funding support via the NSF CAREER award #1944134 and DOE Early Career Research Program award #DE-SC0022915.

*Disclaimer*: Certain equipment, instruments, software, or materials, commercial or non-commercial, are identified in this paper to specify the experimental procedure adequately. Such identification is not intended to imply recommendation or endorsement of any product or service by NIST, nor is it intended to imply that the materials or equipment identified are necessarily the best available for the purpose.

## VI. References

1. Nguyen, L.; Tao, F. F.; Tang, Y.; Doug, J.; Bao, X. J., Understanding Catalyst Surfaces during Catalysis through Near Ambient Pressure X-ray Photoelectron Spectroscopy. *Chemical Reviews* **2019,** *119* (12), 6822-6905.
2. Dupuy, R.; Richter, C.; Winter, B.; Meijer, G.; Schlögl, R.; Bluhm, H., Core level photoelectron spectroscopy of heterogeneous reactions at liquid–vapor interfaces: Current status, challenges, and prospects. *The Journal of Chemical Physics* **2021,** *154* (6).
3. Salmeron, M., From surfaces to interfaces: ambient pressure XPS and beyond. *Topics in Catalysis* **2018,** *61*, 2044-2051.
4. Schnadt, J.; Knudsen, J.; Johansson, N., Present and new frontiers in materials research by ambient pressure x-ray photoelectron spectroscopy. *Journal of Physics: Condensed Matter* **2020,** *32* (41), 413003.
5. Operando NAP-XPS study of plasma-enhanced surface reactions. *SPECS Surface Nano Analysis GmbH* **2023,** *Application Note #230401*.
6. Vandencasteele, N.; Reniers, F., Plasma-modified polymer surfaces: Characterization using XPS. *Journal of Electron Spectroscopy and Related Phenomena* **2010,** *178-179*, 394-408.
7. Naclerio, A. E.; Kidambi, P. R., A Review of Scalable Hexagonal Boron Nitride (h-BN) Synthesis for Present and Future Applications. *Adv Mater* **2022**, e2207374.
8. Moon, S.; Kim, J.; Park, J.; Im, S.; Kim, J.; Hwang, I.; Kim, J. K., Hexagonal Boron Nitride for Next-Generation Photonics and Electronics. *Advanced Materials Interfaces* **2023,** *35* (4), 2204161.
9. Roy, S.; Zhang, X.; Puthirath, A. B.; Meiyazhagan, A.; Bhattacharyya, S.; Rahman, M. M.; Babu, G.; Susarla, S.; Saju, S. K.; Tran, M. K., Structure, properties and applications of two-dimensional hexagonal boron nitride. *Advanced Materials* **2021,** *33* (44), 2101589.
10. Sasama, Y.; Komatsu, K.; Moriyama, S.; Imura, M.; Teraji, T.; Watanabe, K.; Taniguchi, T.; Uchihashi, T.; Takahide, Y., High-mobility diamond field effect transistor with a monocrystalline h-BN gate dielectric. *APL Materials* **2018,** *6* (11).
11. Li, L. H.; Xing, T.; Chen, Y.; Jones, R., Boron Nitride Nanosheets for Metal Protection. *Advanced Materials Interfaces* **2014,** *1* (8).

12. Lo, C.-L.; Helfrecht, B. A.; He, Y.; Guzman, D. M.; Onofrio, N.; Zhang, S.; Weinstein, D.; Strachan, A.; Chen, Z., Opportunities and challenges of 2D materials in back-end-of-line interconnect scaling. *Journal of Applied Physics* **2020,** *128* (8).
13. Uwanno, T.; Hattori, Y.; Taniguchi, T.; Watanabe, K.; Nagashio, K., Fully dry PMMA transfer of graphene on h-BN using a heating/cooling system. *2d Materials* **2015,** *2* (4).
14. Vlassiouk, I.; Smirnov, S.; Puretzky, A.; Olunloyo, O.; Geohegan, D. B.; Dyck, O.; Lupini, A. R.; Unocic, R. R.; Meyer III, H. M.; Xiao, K., Armor for Steel: Facile Synthesis of Hexagonal Boron Nitride Films on Various Substrates. *Adv. Mater. Interfaces* **2023,** 2300704.
15. Kidambi, P. R.; Blume, R.; Kling, J.; Wagner, J. B.; Baehtz, C.; Weatherup, R. S.; Schloegl, R.; Bayer, B. C.; Hofmann, S., In Situ Observations during Chemical Vapor Deposition of Hexagonal Boron Nitride on Polycrystalline Copper. *Chem Mater* **2014,** *26* (22), 6380-6392.
16. Scardamaglia, M.; Boix, V.; D'Acunto, G.; Struzzi, C.; Reckinger, N.; Chen, X.; Shivayogimath, A.; Booth, T.; Knudsen, J., Comparative study of copper oxidation protection with graphene and hexagonal boron nitride. *Carbon* **2021,** *171*, 610-617.
17. Diulus, J. T.; Novotny, Z.; Dongfang, N.; Beckord, J.; Al Hamdani, Y.; Comini, N.; Muntwiler, M.; Hengsberger, M.; Iannuzzi, M.; Osterwalder, J., Towards 2D-confined catalysis on oxide surfaces. *Journal of Physical Chemistry C* **2023,** *Submitted*.
18. Kidambi, P. R.; Chaturvedi, P.; Moehring, N. K., Subatomic species transport through atomically thin membranes: Present and future applications. *Science* **2021,** *374* (6568), eabd7687.
19. Wei, M.; Fu, Q.; Wu, H.; Dong, A.; Bao, X., Hydrogen intercalation of graphene and boron nitride monolayers grown on Pt (111). *Topics in Catalysis* **2016,** *59*, 543-549.
20. He, L.; Wang, H.; Chen, L.; Wang, X.; Xie, H.; Jiang, C.; Li, C.; Elibol, K.; Meyer, J.; Watanabe, K., Isolating hydrogen in hexagonal boron nitride bubbles by a plasma treatment. *Nat. Commun.* **2019,** *10* (1), 2815.
21. Zhang, H.; Feng, P., Controlling bandgap of rippled hexagonal boron nitride membranes via plasma treatment. *ACS Applied Materials Interfaces* **2012,** *4* (1), 30-33.
22. Koswattage, K. R.; Shimoyama, I.; Baba, Y.; Sekiguchi, T.; Nakagawa, K., Selective adsorption of atomic hydrogen on a h-BN thin film. *The Journal of chemical physics* **2011,** *135* (1).
23. Eads, C. N.; Zhong, J.-Q.; Kim, D.; Akter, N.; Chen, Z.; Norton, A. M.; Lee, V.; Kelber, J. A.; Tsapatsis, M.; Boscoboinik, J. A.; Sadowski, J. T.; Zahl, P.; Tong, X.; Stacchiola, D. J.; Head, A. R.; Tenney, S. A., Multi-modal surface analysis of porous films under operando conditions. *AIP Advances* **2020,** *10* (8).
24. Ng, M. L.; Shavorskiy, A.; Rameshan, C.; Mikkelsen, A.; Lundgren, E.; Preobrajenski, A.; Bluhm, H., Reversible Modification of the Structural and Electronic Properties of a Boron Nitride Monolayer by CO Intercalation. *ChemPhysChem* **2015,** *16* (5), 923-927.
25. Greczynski, G.; Hultman, L., X-ray photoelectron spectroscopy: Towards reliable binding energy referencing. *Progress in Materials Science* **2020,** *107*.
26. Trotochaud, L.; Head, A. R.; Karslioglu, O.; Kyhl, L.; Bluhm, H., Ambient pressure photoelectron spectroscopy: Practical considerations and experimental frontiers. *J Phys Condens Matter* **2017,** *29* (5), 053002.
27. Amati, M.; Bonanni, V.; Braglia, L.; Genuzio, F.; Gregoratti, L.; Kiskinova, M.; Kolmakov, A.; Locatelli, A.; Magnano, E.; Matruglio, A., Operando photoelectron emission spectroscopy and microscopy at Elettra soft X-ray beamlines: From model to real functional systems. *Journal of Electron Spectroscopy Related Phenomena* **2022,** *257*, 146902.
28. Masuda, T., Various spectroelectrochemical cells for in situ observation of electrochemical processes at solid–liquid interfaces. *Topics in Catalysis* **2018,** *61*, 2103-2113.